\documentclass[twocolumn,showpacs,prl,groupedaddress,preprintnumbers,amsmath,amssymb]{revtex4}
\usepackage{epsfig}
\usepackage{graphicx}
\usepackage{bm}

\begin{document}

\title{Human group formation in online guilds and offline gangs\\  driven by common team dynamic}
\author
{Neil F. Johnson$^{1}$, Chen Xu$^{2,3}$, Zhenyuan Zhao$^1$, \\
Nicolas Ducheneaut$^4$, Nicholas Yee$^4$, George Tita$^{5}$, Pak Ming Hui$^3$}
\affiliation{$^{1}$Physics Department, University of Miami, FL 33126, U.S.A.\\
$^{2}$Jiangsu Key Laboratory of Thin Films, School of Physical Science and Technology, Soochow University, Suzhou 215006, China\\
$^{3}$Department of Physics, The Chinese University of Hong Kong, Shatin, Hong Kong, China\\
$^{4}$Palo Alto Research Center,
Palo Alto, CA 94304, U.S.A.\\
$^{5}$Department of Criminology, Law
and Society, University of California, Irvine, CA 92697, U.S.A.}

\date{\today}


\begin{abstract}
{Quantifying human group dynamics represents a unique challenge.
Unlike animals and other biological systems, humans form groups in
both real (offline) and virtual (online) spaces -- from
potentially dangerous street gangs populated mostly by disaffected
male youths, through to the massive global guilds in online
role-playing games for which membership currently exceeds tens of
millions of people from all possible backgrounds, age-groups and
genders. We have compiled and analyzed data for these two
seemingly unrelated offline and online human activities, and have
uncovered an unexpected quantitative link between them. Although
their overall dynamics differ visibly, we find that a common
team-based model can accurately reproduce the quantitative
features of each simply by adjusting the average tolerance level
and attribute range for each population. By contrast, we find no
evidence to support a version of the model based on
like-seeking-like (i.e. kinship or `homophily').}
\end{abstract}


\maketitle




\section{I. INTRODUCTION}
Quantifying the group dynamics of living objects is a fundamental
challenge across the sciences
\cite{mike,iain,levin,levin2,epstein,cederman,humans,Onnela,roger,barabasi,uzzi,palla}.
Humans represent a particularly difficult case to analyze, since
their groups can be formed in both real (offline) and virtual
(online) spaces. Such fascinating sociological challenges have
attracted the attention of many physicists in recent years under
the heading of Econophysics and Sociophysics
\cite{Onnela,roger,barabasi,uzzi,palla,galam}. Indeed, the
Econophysics website provides a rapidly increasing repertoire of
such investigations \cite{Econophysics}.

Massively multiplayer online games typically allow individuals to
spontaneously form, join or leave a formal group called a guild
\cite{nic1,nic2}. The design of the game encourages players to
form such groups by making the most rewarding quests (i.e.,
missions) too difficult to accomplish alone. Millions of people
worldwide log on to the world's largest online game (World of
Warcraft (WoW)) for the equivalent of several days every week.
Indeed, online games are one of the largest collective human
activities on the planet and hence of interest from the
perspectives of global commerce \cite{IBM}, security \cite{terror}
and even epidemiology \cite{epid}. A seemingly unrelated social
phenomenon which is also of great concern, is urban gangs. Urban
gangs have been gaining in popularity among young people both
nationally and internationally \cite{Justice,bbc,george}. There
are obvious differences in the settings and history of online
guilds and offline gangs, however the empirical datasets that we
have compiled enable us to perform a unique comparative study of
their respective grouping dynamics \cite{nic1,nic2,george}.

Studies of the formation and evolution of groups have long
occupied a central position within the sociological and
organizational theory literatures, particularly in terms of
understanding how individual level characteristics (e.g.,
demographics, skill sets) shape group dynamics
\cite{levin,levin2,lazer,barabasi,ibm2,kreps,uzzi,roger, palla,
mike,iain,epstein,cederman,mayhew,ruef}. Proponents of homophily
tend to argue that individuals choose to participate in groups
that minimize within-group heterogeniety, since sameness
facilitates communication and reduces potential conflict
\cite{levin2,lazarsfeld,blau,McPherson1,McPherson2}. With respect
to stability, previous research has suggested that members of
groups that are most unlike the other members of the group are
also more likely to exit the group \cite{Popielarz}. By contrast,
some researchers suggested that rather than minimize diversity
among members, members might instead join groups that maximize the
diversity of skills in the group (team) \cite{Thibaut, Hinds}
since a wider skill set might be more efficient in meeting
particular goals \cite{uzzi,Thibaut, Hinds}.

In this paper, we analyze data obtained from street gangs in the
offline, real world \cite{Justice,bbc,george} and Internet guilds
in virtual online worlds within massively multiplayer online
role-playing games \cite{nic1,nic2,IBM}.  We develop and employ a
physically motivated model to analyze these two high profile, yet
seemingly unrelated, human activities. The underlying datasets
were obtained from online WoW guilds \cite{nic1,nic2} and urban
street gangs in Long Beach, California \cite{george}. They have
been separately compiled by members of our team over the past few
years through a combination of field-work and data compilation,
and are believed to be state-of-the-art datasets for each
respective system. As a result of our analysis, we uncover
evidence of a quantitative link between the collective dynamics in
these two systems. Although the observable group-size
distributions are very different, we find that a common
microscopic mechanism can reproduce the observed grouping data for
each, simply by adjusting the populations' average attribute
property. In particular, we find that the evolution of gang-like
groups in the real and virtual world can be explained using the
same team-based group formation mechanism. In contrast to the
quantitative success of our team-based model, we find that a
homophilic version of the model fails.  Our findings thus provide
quantitative evidence that online guilds and offline gangs are
both driven by team-formation considerations, rather than
like-seeking-like. Interestingly, each server's Internet Protocol
(IP) address seems to play an equivalent role to a gang ethnicity.
Given the current public concern regarding the social consequences
of intensive Internet game-playing, and separately the current
rise in street gangs \cite{Justice,bbc,george}, we hope that the
present findings help contribute to the debate by setting these
systems on a common footing.

The plan of the paper is as follows: Sec. II gives the main
empirical results that are to be modelled.  Section III gives the
key ideas and a detailed description of our self-organized team
formation model.  The main results comparing the cumulative group
size distributions from our model and real data for both WoW
guilds and LA gangs are also presented, in order to establish the
validity of our model.  Section IV gives further analysis of the
non-cumulative WoW guild size distributions for separate servers,
as well as the group size distributions for LA gangs of different
ethnic groups.  These results represent a more stringent test of
our model.  Section V defines the kinship model and demonstrates
its inadequacy.  Finally, Sec. VI provides the conclusions and
discusses the implications. Note that our philosophy throughout this work was to see if we could identify a minimal model which is consistent with the empirical observations from two very different human grouping activities -- one offline and one online. More complex models can of course be built, and may even agree better with particular portions of the empirical data. Likewise, we cannot prove that the model that we propose is strictly {\em the} minimal model. However, we have explored many similar models, and the one that we present seems the most reasonable, least complicated and provides the best empirical fit.

\begin{figure}
\includegraphics[width=6.0 cm]{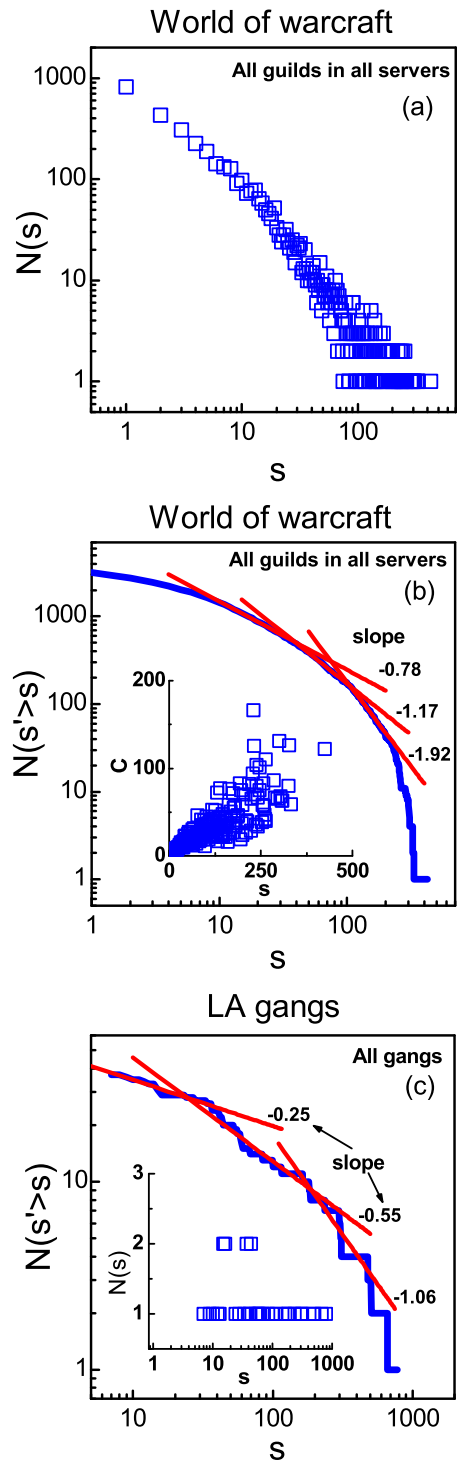}
\caption{(Color Online) Internet guilds and street gangs. (a)
Empirical data from World of Warcraft on all servers. (b)
Cumulative distribution differs significantly from a power-law.
Inset shows the averaged churn $C$ of the guilds. (c) Cumulative
distribution for Long Beach (i.e. `LA') gangs. Inset shows the
underlying discrete distribution.}
\end{figure}

\section{II. Main Empirical Results}

As explained in Ref. \cite{nic1}, World of Warcraft (WoW) is a
massively multiplayer online role-playing game, where players
control a character avatar within a virtual world, exploring the
landscape, completing quests and interacting with other players.
To enter the game, the player must select a particular realm (or
server), each of which acts as an individual copy of the game
world \cite{nic1}. Within the game itself, players can group
themselves into guilds which may offer an advantage when tackling
certain challenges within the game. Guilds are self-organized
groups whose size, composition and lifetime are not fixed or
pre-determined. Instead, the population tends to self-organize
itself into an evolving ecology of such guild groups -- and their
exact size and composition can be accessed at any time from the
electronic records stored on all the participating servers. By
contrast, most studies of social group behavior in the real world
are plagued by a lack of dynamical information about group size
and membership, and the practical restrictions to small samples
(as opposed to several million WoW players) and specific
geographical locations (as opposed to the global Internet). This
explains the attraction of such online games for studying
collective human behavior.  Within the WoW game itself, each
player can be regarded as having certain attributes. While such an
attribute list in practice may be an extended vector, we take the
minimal model approach of expressing this attribute in terms of
the measures of a single variable (see later). The decision to
leave or join a guild (i.e. group) does not lie solely at the
discretion of the individual player involved -- instead, the
player and guild in question must find each other to be mutually
acceptable.

Figures 1(a) and 1(b) show the guild size distribution $N(s)$ and
cumulative guild size distribution $N(s'>s)$ for a typical
one-month period within the full WoW dataset. The full WoW dataset
itself was collected from three different servers -- each
representing a different game environment -- between June 2005 and
December 2005, and is representative of the entire game's recent
history. There are 76686 agents involved in a total of 3992 guilds
spread across three servers: S1, S2, and S3. The cumulative
distributions for the separate servers S1, S2 and S3 will be shown
later (see Fig.~6(a)). All three servers are based in the US and
were selected at random, with the servers' identities anonymized
to preserve players' privacy. The vertical axis $N(s)$ is the
number of guilds of size $s$. Data is shown using October 2005 as
a representative month, however other months show similar behavior
as demonstrated in a later section. Interestingly, the
distribution is neither a Gaussian nor a power-law. Figure 1(b)
confirms that if we were to insist on power-law behavior, the
supposedly constant slope in $N(s'>s)$ would vary unacceptably.
The inset in Fig.~1(b) shows the quantity called the averaged
churn $C$ versus the guild size $s$, where $C$ describes the
monthly guild dynamics as follows: The membership of a guild is
recorded at the beginning and end of each month, with the churn
being the number of players who were members at the beginning of
the month but who then left during that month. For guilds which
have the same size at the beginning of the month, we then average
over the churn values and call this averaged quantity $C$. We have
checked across different months, and have also looked at different
measures, in order to convince ourselves that the data in Fig.~1
are typical of the WoW data. Figure 1(c) shows our empirical data
for the 5214 members of street gangs in Long Beach, California
just outside of Los Angeles. The data are shown for June 2005, but
again other months show similar behavior. For convenience, we
label these as `LA gangs'. All gangs are included irrespective of
their ethnicity (e.g. Latino). The number of real gangs is much
smaller than the number of guilds in WoW. $N(s'>s)$ for gangs is
not smooth -- nor is it a power-law with a well-defined slope, as
shown explicitly in Fig.~1(c).

\begin{figure}
\includegraphics[width=6.0 cm]{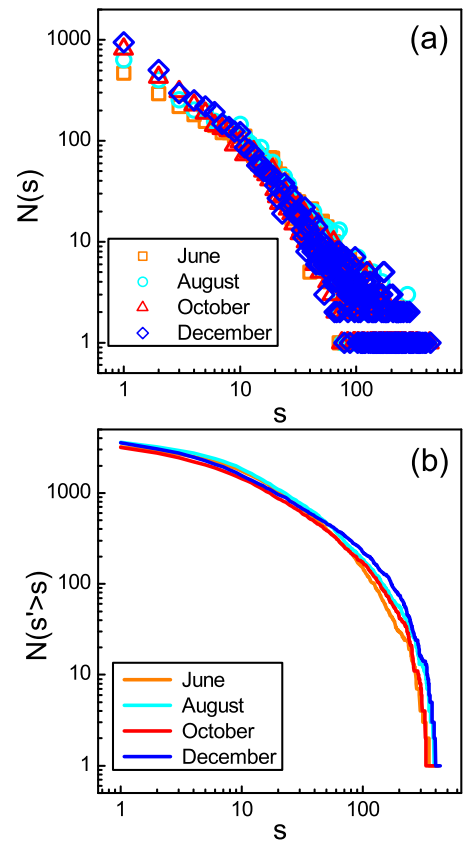}
\caption{(Color Online) (a) WoW guild size distributions $N(s)$
for the months June, August, October, and December 2005. The total
numbers of players in these months are 80183, 93127, 76686, and
93322, respectively. (b) The cumulative guild size distributions
$N(s'>s)$ for each of the four months.} \label{fig:SIfigureS1}
\end{figure}

\begin{figure}
\includegraphics[width=7.0 cm]{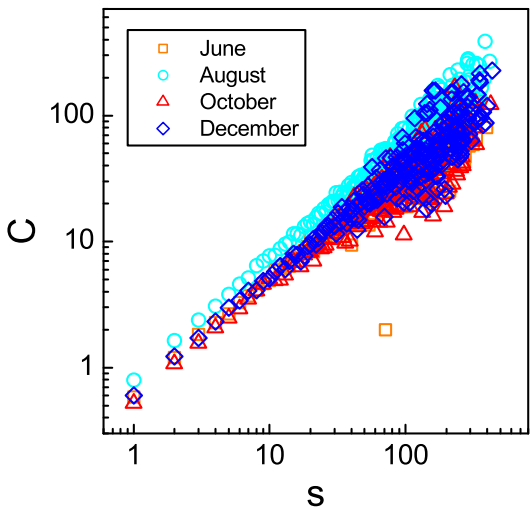}
\caption{(Color Online) The average churn $C$ (as defined in text)
as a function of guild size in the WoW dataset on a log-log plot,
treating the data in all three servers collectively. Data are
shown for the months June, August, October, and December 2005.
Note the inset of Fig.1(b) shows the same data, but on a linear
scale and for October only.} \label{fig:SIfigureS2}
\end{figure}

\begin{figure*}
\includegraphics[width=15.0 cm]{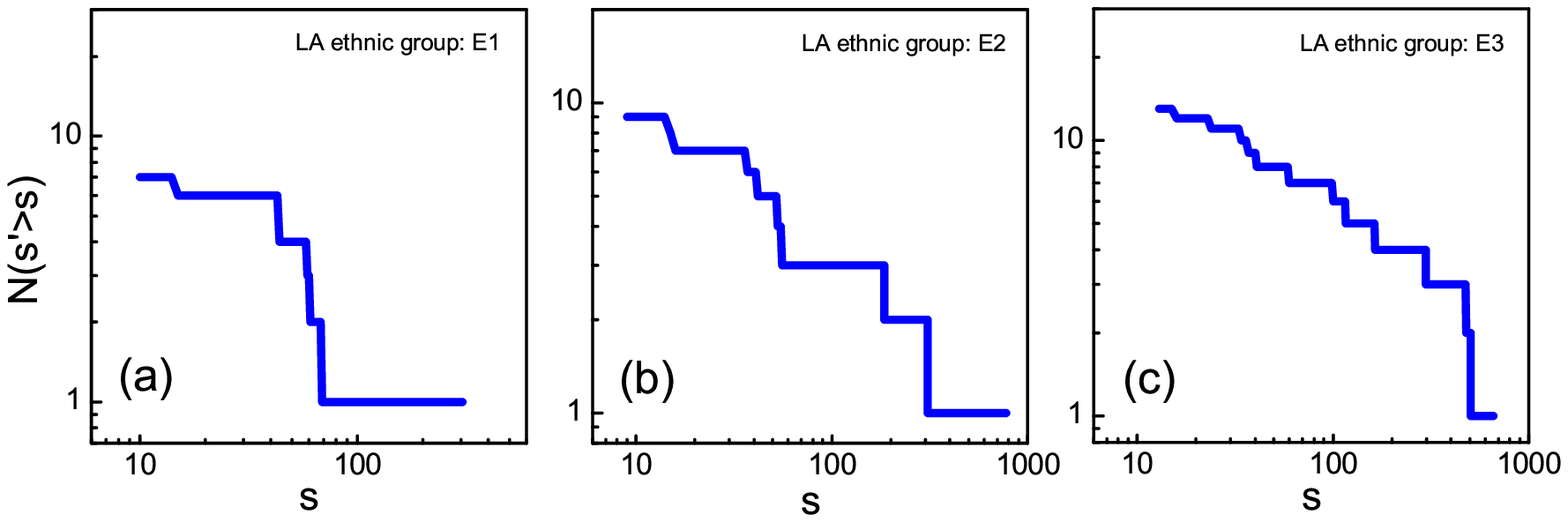}
\caption{(Color Online) The cumulative gang size distribution
$N(s'>s)$ for LA gangs of three main ethnic groups. (a) Cumulative
gang size distribution for gangs with ethnicity E1. The total
membership is $N=608$. (b) Ethnicity E2 with total membership
$N=1504$.  (c) Ethnicity E3 with total membership $N=2552$.}
\label{fig:SIfigureS3}
\end{figure*}

\subsection{A. WoW: Monthly Guild Size Distributions and Churn}
\begin{figure*}
\includegraphics[width=13.0 cm]{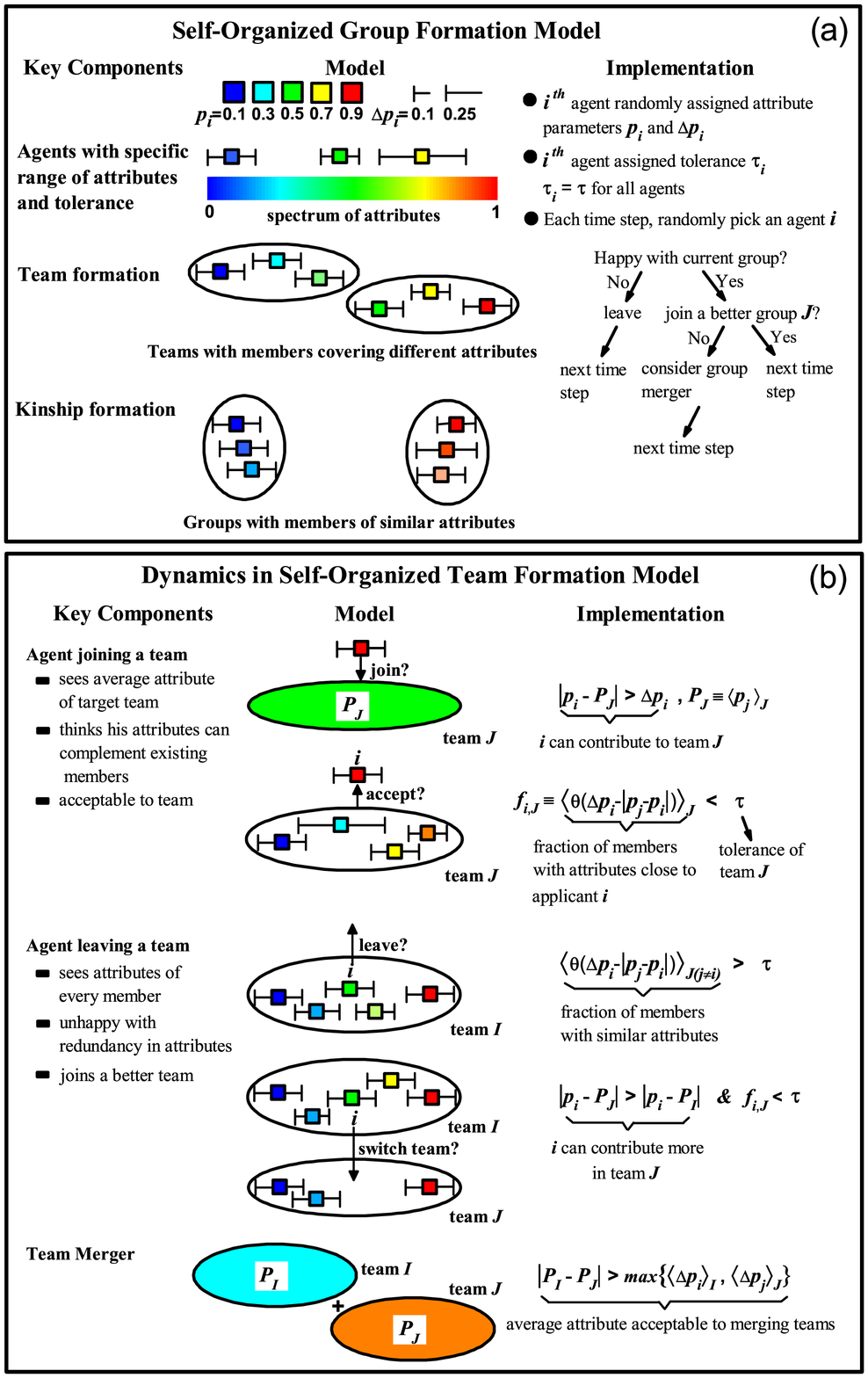}
\caption{(Color Online) Our generic model of group dynamics. (a)
The basic model setup, without yet specifying the criterion that
an agent uses when seeking to join or leave a group. Two possible
extremes are the team-formation model shown in Fig.~5(b), where an
agent seeks a group with a suitable niche in $p$-space, and the
kinship model (not shown) where an agent seeks a group having
members with a similar $p$-value.  Details of the implementation
and specific rule-sets are discussed in Sec.III.}
\end{figure*}

To demonstrate that the form of the distribution in October 2005
is typical of the WoW data, we also analyzed the data for all the
remaining months. For each month, we repeat the same exercise of
counting the guilds and their sizes for each server. Here, we show
the data for several additional months (i.e. June, August and
December 2005) as well as October 2005. The empirical data shows
that the number of players in each month was 80183 (June), 93127
(August), 76686 (October), and 93322 (December). Figure
\ref{fig:SIfigureS1} shows $N(s)$ and $N(s'>s)$ for these four
months. The distributions for different months behave in a similar
way. The results indicate that the guild size distribution
measured at any time during the data collection process,
represents a general property of the game during the entire data
collection window.

In the inset of Fig.~1(b), we showed the values of $C$ for all the
guilds in the three servers (S1, S2, S3) for October 2005. Here,
the data of $C$ for June, August, and December 2005 are shown in
Fig.~\ref{fig:SIfigureS2}. The data indicate that the behavior of
$C$ versus guild size is almost the same for every month. Thus,
the behavior $C \sim s$ is a general feature of the WoW data. We
have also analyzed the data for separate servers, and the behavior
is again nearly the same. Note that there are necessarily fewer
data points for a single server, hence it is more convenient to
show the results corresponding to all servers bundled together.
Later, we will compare results of $N(s)$ as obtained by our
team-formation model with data of separate servers (see
Fig.~\ref{fig:SIfigureS4}).

\subsection{B. LA gangs: Different ethnic groups}
Our dataset on LA gangs collected in June 2005 consists of the
sizes and the ethnicity of the gangs.  Putting all the data
together, there are a total of 5214 members. The cumulative
distribution $N(s'>s)$ was shown in Fig.~1(c). The distribution
shows a similar shape as for the WoW cumulative guild size
distribution. From the information on the ethnicity of the gangs,
there are three main ethnic groups that one can identify. For
privacy reasons, we label these groups as E1, E2, and E3, with
membership 608, 1504, 2552, respectively. Figure
\ref{fig:SIfigureS3} shows the cumulative gang size distributions
$N(s'>s)$ for the three major ethnic groups.  For each of these
ethnic groups, the number of gangs is very small (around 10).  For
this reason, $N(s'>s)$ shows step-like behavior. Comparing with
WoW data, the total number of gangs and the number of members in
the LA gang data are both much smaller than the corresponding
numbers in WoW.  In a later section, we will compare the results
for our team-formation model with these data of different
ethnicity groups (see Fig.~\ref{fig:SIfigureS5}).

\section{III. SELF-ORGANIZED TEAM FORMATION MODEL AND MAIN MODELLING RESULTS}

We now introduce the key ideas within our model,  describe its
details, and show that it reproduces accurately the quantitative
features of the empirical data.  As an overview, Fig.~5(a) shows
our generic model of self-organized group formation which acts as
the core setup for implementing specific rule-sets for joining and
leaving a group -- for example, team-formation (see Fig.~5(b)) or
homophilic kinship.  Our generic model (Fig.~5(a)) creates a
heterogeneous population by assigning an attribute $p_{i}$ to each
person (i.e. agent) $i$. Since people may have a range of
attributes, we assign each agent a spread $\Delta p_{i}$ around
$p_{i}$.  With the goal of building a minimal model, we choose
each $p_{i}$ to be a single number chosen randomly from a uniform
distribution between zero and one.  More complicated models can of
course be built by assigning, for example, an array of numbers to
describe the attributes of a person -- however we again stress
that we are seeking a minimal model in the present work.  The
values of $\Delta p_{i}$'s are random numbers drawn from a
single-peaked distribution with mean $\langle \Delta p_i\rangle$
and spread (i.e. standard deviation) $\sigma_{\Delta p}$.  The
$\Delta p_{i}$ values are shown in Fig.~5 as horizontal bars
around the corresponding color-coded $p_{i}$ value. We then assign
a tolerance to every agent -- for simplicity, we choose the same
value $\tau$ for each agent. The tree on the right-hand side of
Fig.~5(a) applies to both team- formation and kinship versions. In
the team-formation version, the group contains agents with
complementary attributes (i.e. a team) while in the kinship
version a group contains agents with similar attributes (i.e. like
with like).

The model can be constructed without considering a particular
context.  It could represent players in WoW, members in gangs,
employees in companies, etc.  Figure~5(b) describes what happens
in one timestep in the team-formation implementation of Fig.~5(a),
both schematically and mathematically.  The kinship model,
described later, essentially corresponds to an opposite set of
add-on rules to the team-formation model.  The team-formation
model, as we shall see, works better for the empirical data and we
will focus on it in this section.  We will use the words `team'
and `group' interchangeably in the following discussion. However
we emphasize that for the portions of the following discussion
concerning Fig.~5(a), the word `team' can be replaced by `group'
since the statements apply equally to the team-formation model and
the kinship model.

{\em Parameters --} Consider a population of $N$ agents or
players. The attributes of an agent $i$ are described by a set of
numbers $(p_{i}, \Delta p_{i}, \tau_{i})$, where $p_{i}$ describes
the $i$th-agent's mean attribute. $\Delta p_{i}$ describes the
$i$-th player's range of attributes around $p_{i}$, or
equivalently a breadth of skills around the mean skill.  The value
of $\Delta p_{i}$ is independent of the value of $p_{i}$. Here,
$\tau_{i}$ is a parameter that describes the tolerance of an agent
in deciding whether to leave a group, after he compares how close
his attributes are to the members of the group. In the present
model, we have not included the possible evolution of attributes,
although this is an interesting problem for future studies.

{\em Initialization --} Initially, each agent is randomly assigned
his attribute parameter $p_{i}$, the value of which is chosen
randomly from a uniform distribution between $0$ and $1$.  The
agents' $\Delta p_{i}$'s are assumed to follow a Gaussian
distribution characterized by a mean $\langle \Delta p_i\rangle$
and standard deviation $\sigma_{\Delta p}$.  Each agent is then
assigned a value of $\Delta p_{i}$ from this Gaussian
distribution. With $p_{i}$ and $\Delta p_{i}$, the agent $i$
covers the attributes $p_{i} \pm \Delta p_{i}$, for attributes
characterized by the range between $0$ and $1$. The coverage of
attributes is not allowed to go below $0$ or above $1$, i.e., when
$p_{i}+\Delta p_{i} > 1$, the upper bound is set at $1$ and when
$p_{i} - \Delta p_{i} < 0$, the lower bound is set at $0$. For
simplicity, the values of $\tau_{i}$ are taken to be the same for
all agents, i.e., $\tau_{i} = \tau$ for all agents. The total
number of agents in the system $N$ can be easily taken from the
real data.  Thus, the model is completely characterized by four
physically meaningful parameters: $N$, $\langle \Delta
p_i\rangle$, $\sigma_{\Delta p}$ and $\tau$.

{\em Key Ideas and Model Implementation --} In each timestep, an
agent $i$ is randomly picked. The attachment of the agent $i$ to a
group then follows the rules below.

\begin{itemize}

\item[(a)] {\bf For a single agent joining a team --} This step is
imposed when the agent $i$ being picked is an isolated agent. In
this case, another agent $j$ is randomly picked.  The agent $j$
belongs to a team labelled $J$ with $n_{J}$ members.  Note that
$n_{J}=1$ if $j$ is an isolated agent.  The key idea is that it is
a two-way consideration when an agent $i$ wants to join a team
$J$: the agent must find a team to which his attributes could
contribute, and that team must in turn find the agent's attributes
acceptable. Moreover, the agent can only see the average
attributes of the team to which he is applying. In other words,
when joining a team, an agent will be guided by general
information about the team (i.e. the average attribute of the
team) rather than detailed information about all its members. This
mimics the fact that an outsider cannot be expected to be aware of
all the details of a team's members before joining, since such
knowledge can generally only be gained after being a member of
that team. Once inside the team, this information can then be
gained either through direct access to insider knowledge, or
simply through osmosis.

An agent $i$ therefore assesses a team $J$ which he considers
joining, by looking at the average attribute $P_{J}$ of that team:
\begin{equation}
P_{J} = \frac{1}{n_{J}} \sum_{k \in J} p_{k},
\end{equation}
where the sum is over all members of team $J$.  The agent $i$ will
find the team suitable if his attributes complement those of the
existing members.  Therefore, if his attributes are too close to that
of the existing members of the team, he feels that he could not
contribute much and he will not join the team.  The condition that
the agent $i$ finds the team $J$ acceptable can thus be modelled
by $|p_{i} - P_{J}| > \Delta p_{i}$.

For the team $J$, it will consider whether to enroll agent $i$ as
a new member.  As an applicant to the team $J$, the team will know
the range of attributes that agent $i$ could cover and then assess the
potential contribution of agent $i$ to the team.  This can be
measured by counting the number of existing members with attributes in
the range of agent $i$, normalized by the team size
$n_{J}$.  Thus, we define $f_{i,J}$ as
\begin{equation}
f_{i,J} = \frac{1}{n_{J}} \sum_{j \in J} \theta(\Delta p_{i} -
|p_{j}-p_{i}|),
\end{equation}
where the sum is over all members in team $J$ and $\theta(x)$ is
the Heaviside function, i.e., $\theta(x) = 1$ for $x > 0$ and
$\theta(x) = 0$ otherwise.  In deciding whether to accept a new
member, we define a team's tolerance by averaging the individual
tolerance of its members, i.e.,
\begin{equation}
\tau_{J} = \frac{1}{n_{J}} \sum_{j \in J} \tau_{j}.
\end{equation}
For $\tau_{j} = \tau$ for all agents, $\tau_{J} = \tau$.  Note
that $f_{i,J}$ is a quantity less than unity.  If $f_{i,J}$ is
large, many existing members in team $J$ have attributes that are
close to that of agent $i$ and thus the team tends not to accept agent
$i$ as a new member due to redundance in attributes.  Thus, the
condition that the team $J$ will accept agent $i$ as a new member
is $f_{i,J} < \tau_{J}$.

Considering joining a team requires two-way consideration, the
criteria for an agent $i$ joining a team $J$ are: $|p_{i} - P_{J}|
> \Delta p_{i}$ and $f_{i,J} < \tau_{J}$.

\item[(b)] {\bf For an agent leaving a group, finding a better
group, or for groups merging --} This step is imposed when the agent
$i$ being picked belongs to a group labelled $I$ with $n_{I}$
($n_{I}>1$) members. The following attempts are implemented {\bf
in sequence}.
\end{itemize}

\begin{itemize}
\item[(i)] {\bf Agent $i$ decides whether he can tolerate the
team} -- After being a member of team $I$ for a while, the agent
$i$ has the chance to explore the microscopic details (individual
attributes) of the team members.  The key idea is that if he finds
that there are many members with similar attributes to his, he
will leave.  To decide whether he can tolerate the team, he looks
at the fraction $f_{i}$ of members in the team with attributes
within his range of coverage, i.e.,
\begin{equation}
f_{i} = \frac{1}{n_{I}-1} \sum_{k \in I (k\neq i)} \theta(\Delta
p_{i} - |p_{k}-p_{i}|),
\end{equation}
where the sum is over all the agents in the team $I$ except the
agent $i$ himself.  Note that $0 \leq f_{i} \leq 1$. If $f_{i}$ is
close to $1$, then there are too many members with similar attributes
and the agent $i$ will have a higher tendency to leave.  If $f_{i}
> \tau_{i}$, the $i$-th agent cannot tolerate the team any more
and he {\em leaves} the group to become an isolated agent.  If
this happens, the timestep ends.

\item[(ii)] Another key idea is team switching.  If the agent $i$
finds that he can tolerate the team, it does not necessarily mean
that he is very happy with the team. He will try to find a better
(more suitable) team to join. An agent $j$, who belongs to a group
$J$, is randomly picked.  The agent $i$ will then compare whether
the current team $I$ or the team $J$ is more suitable for him.  He
intends to join team $J$ if $|p_{i} - P_{J}|
> |p_{i} - P_{I}|$.  This criterion implies that the agent $i$
finds that he can contribute more in team $J$ than in
team $I$.  Whether team $J$ would accept agent $i$ as a new member
is again determined by the criterion $f_{i,J} < \tau_{J}$, as in
step (a).  Thus, the criteria for agent $i$ to switch from team
$I$ to team $J$ successfully are $|p_{i} - P_{J}|
> |p_{i} - P_{I}|$ and $f_{i,J} < \tau_{J}$.  If there is group
switching, the timestep ends.  We remark that the steps (b)(i)
and (ii) are similar to job hunting. If the job is too bad, then
one will simply quit without finding a new job. This is reflected
in (b)(i). However, even if the job is acceptable, one tries to look
for a better job. In job hunting, it is a two-way process: The
employer is looking for someone who can cover the weaker aspects
or services in a company, and the employee is looking for a better
place. This is reflected in (b)(ii).

\item[(iii)] The next key idea is to allow for team mergers. If
nothing actually happened in (i) and (ii), i.e., the $i$-th agent
does not leave the team $I$, either because he is happy or because
team switching does not work, we consider the possibility of
allowing two teams to merge.  Team $I$ to which agent $i$ belongs,
merges with another team $J$ under the criterion $|P_{I} - P_{J}|
> \Delta P_{I}$, where $\Delta P_{I} = (1/n_{I}) \sum_{i \in I}
\Delta p_{i}$.  Similarly, team $J$ considers merging with
team $I$ under the criterion $|P_{J} - P_{I}|
> \Delta P_{J}$.  That is to say, if $|P_{I} - P_{J}| >
max(\Delta P_{I}, \Delta P_{J})$, then teams $I$ and $J$ merge to
form a bigger team.  Note that there are two ways to implement
mergers.  The team $J$ could be the same team that the $j$-th
agent belonged to in procedure (ii) above, or a new agent $j$ can
be picked randomly when mergers are considered.  Results are
nearly identical for the two ways.
\end{itemize}

To summarize, the key ingredients in our team-formation model are:
(i) Teams tend to recruit members to cover a spectrum of
attributes; (ii) agent joins a team by assessing his potential
contribution to the team; (iii) agent joining a team only sees an
average of the attributes of a team; (iv) team accepts new member
by assessing his potential contribution; (v) agent leaves a team
when there are many members with similar attributes; (vi) agent
always looks for better teams where he could contribute more; and
(vii) team tends to expand by mergers when its membership becomes
stable. Each of these ingredients seems reasonable based on our
common knowledge of how people behave in team situations. We
remark that this set of rules allowed us to produce results that
are similar to the empirical observations for both the WoW and
gangs data.  If more data become available in order to put further
constraints on the model, or if we only want to model a particular
subset of the behaviors arising in the full dataset, then the
model's rules may either require further elaboration or be further
simplified. For example, ingredient (vi) is needed to model the
averaged churn $C$ in WoW data while ingredient (vii) is needed to
get at the proper sizes of the bigger groups.

\begin{figure}
\includegraphics[width=8.0 cm]{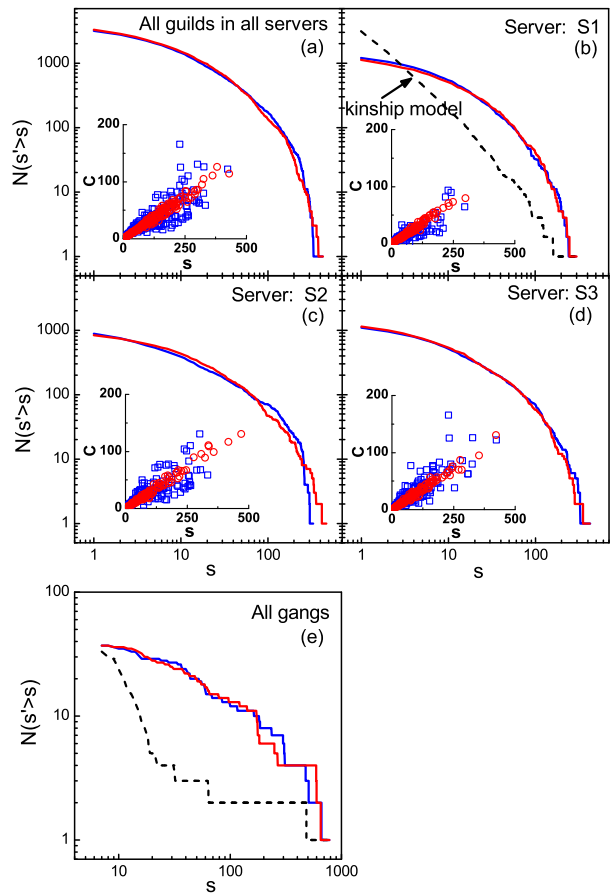}
\caption{(Color Online) Empirical data and model comparison for
(a)-(d) World of Warcraft and (e) LA gangs. Empirical data are
dark blue, and the team-formation model from Fig.~5(b) is in red.
The kinship model (light blue) produces a poor fit in both cases.}
\end{figure}
Figure~6 highlights the main modelling results.  Figure~6(a) shows
that excellent agreement is obtained across the entire range of
observed group sizes for $N(s'>s)$, between the empirical WoW
guild data from Fig.~1(b) (dark blue) and the team-formation model
(red) of Fig.~5.  Throughout this paper, the model(s) is
implemented with the observed number of agents as an input. Here
$\tau= 0.69$, $\langle \Delta p_i\rangle = 0.16$ and
$\sigma_{\Delta p} = 0.022$, but we stress that good agreement can
be obtained across a reasonably wide range of parameter choices.
The remaining panels (Fig.~6(b)-(d)) show the data separated by
server. The parameter values used are within 10$\%$ of those
quoted above.  To calculate $C$ in the model, we record the
membership for each guild in a run during $0.7$ Monte Carlo
timesteps (after a transient of 1000 Monte Carlo steps).  A Monte
Carlo time step is the duration over which each agent has, on
average, been chosen once for carrying out the dynamics in the
model, i.e. each agent has been given a chance to join or leave a
group. We have tried different time windows so as to obtain the
averaged churn $C$ in WoW data and found that $0.7$ Monte Carlo
timesteps happen to give particularly good agreement. One might
interpret this by claiming that the timescale over which $70\%$ of
agents  have a chance to carry out a dynamical update process,
represents the real-world timescale for the churn process --
however, it is very hard to associate timescales in simulations
with specific timescales in the real world. A similar process is
followed for the LA gangs in Fig.~6(e). In an analogous way to the
breakdown by computer server in Fig.~6(a)-(d), one can break down
the LA gang data by ethnicity. The fit by gang ethnicity (see
Sec.IV later) is good even though the numbers are much smaller
than WoW and hence more prone to noise. This surprising connection
between ethnicity and server is consistent with the fact that it
is essentially impossible to change one's real-world ethnicity or
virtual-world server (unless a large fee is paid to WoW in the
latter case, and even then it is an irreversible process). It is
also intriguing that the best-fit model parameter values are so
similar across WoW servers, and across gang ethnicities. This
suggests a quasi-universal behavior in terms of the {\em way} in
which people form gang-like groups online and offline. The small
observed server-dependences (and ethnicity-dependences) can be
explained by players on different servers (and gang members of
different ethnicities) perceiving their environments differently,
and hence adopting slightly different tolerances. Our
team-formation model thus manages to capture all the features of
the empirical gang and guild dynamics, including the approximately
linear increase of the averaged churn $C$ with guild size in WoW.
By contrast, the kinship (i.e. homophilic) version (see Sec.V
later) of the model does {\em not} reproduce the empirical results
of either WoW or the gangs, even qualitatively, as demonstrated by
the light blue curves in Fig.~6.

\section{IV. Further Analysis: $N(s)$ WoW Guild Size Distribution
and $N(s'>s)$ for LA gangs of different ethnic groups}

The agreement of our model with WoW data in $N(s'>s)$ for all and
individual servers (Fig.~6(a)-(d)) can be further illustrated by
comparing the underlying, i.e., non-cumulative, distribution for
the guild size distribution $N(s)$.  Since $N(s)$ is less smooth
and thus more noisy than the cumulative distribution $N(s'>s)$, we
are actually executing a more stringent test of the model by
carrying out the team-formation model comparison based on $N(s)$
instead of $N(s'>s)$.

\begin{figure}
\includegraphics[width=8.0 cm]{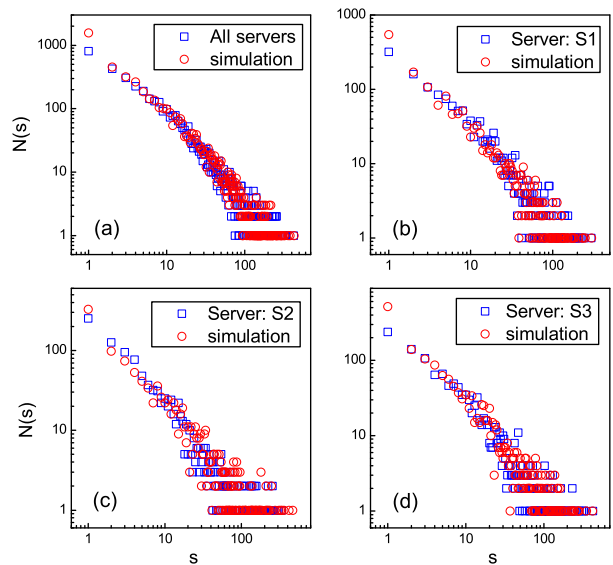}
\caption{(Color Online) The WoW guild size distribution $N(s)$ in
October 2005. (a) Guild size distribution treating all servers
collectively. The parameters used for team formation are
$N=76686$, $\langle \Delta p_i\rangle = 0.160$, $\sigma_{\Delta p}
= 0.022$, and $\tau = 0.69$. (b) Guild size distribution of server
S1. The parameters used for team formation simulation are
$N=24033$, $\langle \Delta p_i\rangle = 0.160$, $\sigma_{\Delta p}
= 0.020$, and $\tau = 0.67$. (c) Guild size distribution of server
S2. The parameters used for team formation simulation are
$N=24477$, $\langle \Delta p_i\rangle = 0.160$, $\sigma_{\Delta p}
= 0.025$, and $\tau = 0.75$. (d) Guild size distribution of server
S3. The parameters used for team formation simulation are
$N=28176$, $\langle \Delta p_i\rangle = 0.161$, $\sigma_{\Delta p}
= 0.020$, and $\tau = 0.70$. Each simulation result is obtained
from one particular run of the team-formation model.  Note that
the parameters for different servers are very similar.}
\label{fig:SIfigureS4}
\end{figure}

From the WoW dataset, we count the number of players in all the
guilds in each server, and also the total number of players in all
servers. In each case, we take the number of players as input for
$N$ and run our team-formation model.  By adjusting the parameters
$\langle \Delta p_i\rangle$, $\sigma_{\Delta p}$ and $\tau$ in the
model, we obtained the guild size distributions $N(s)$ for each of
the three separate servers and for the three servers collectively.
Figure \ref{fig:SIfigureS4} shows the $N(s)$ for our model
obtained from one run in each of these cases, together with the
distribution obtained from the data.  The parameters are given in
the figure caption.  The results from the team-formation model
capture the essential features in the WoW guild size
distributions.

From the parameters for each of the servers, it can be seen that
they are very similar but not identical.  This indicates that
while the behavior of the players in different servers are not too
different, there are slight differences indicating some kind of
special characteristic of a server or game environment. We will
see that similar features also appear in the LA gang data, when
treating ethnicity separately.  To the extent to which the server
identity mimics an ethnicity, this seems to open up some deeper
sociological questions which can be explored in future research on
guilds and gangs.

From our attempts in modelling the real data, we now make a few
comments on the model as related to the key features in real data:
Step (a) (see Sec.III) that sets the criteria for an agent to join
a team and a team to accept a new member is the essence of the
team-formation model.  This is essential in getting the {\em
shape} of $N(s'>s)$. We observed that the shape of $N(s'>s)$, and
thus $N(s)$, is more sensitive to the parameter $\tau$. In the WoW
data, there is a quantity called $C$. In order to get reasonable
values for $C$, a mechanism is required for agents to leave a team
or to switch teams readily. Steps (b)(i) and (b)(ii) serve to
provide such a mechanism.  In order to get at the largest size of
the guilds in real data, we need a mechanism for guilds to merge.
Step (b)(iii) serves this purpose.

If we were to focus {\em only} on fitting the guild-size or
gang-size distributions, and hence decided not to care about
simultaneously fitting the churn $C$ in the WoW data, we could
construct even simpler versions of our model and yet still obtain
group-size distributions similar to the real data. For example, a
model with slower team switching and more static groups can be
used to get at $N(s'>s)$ similar to real data. However, with our
present team-formation model we have managed to fit these size
distributions {\em and} account for the churn. One implication of
our work is therefore that previous grouping models which have
been proposed to explain time-averaged group sizes in real data
{\em without} churn, should be re-examined once such churn data
becomes available. Fitting churn as well as the group-size
distribution presents a stringent challenge which relatively few
candidate models will survive. Performing studies analogous to our
present one, would therefore be a very useful way of reducing the
number of competing models. Likewise our own extensive
experimentation indicates that it would be very hard to identify
an alternative model to our team-formation one, in which equally
high quantitative accuracy was obtained and yet the structure
and/or set of microscopic rules were fundamentally different. This
gives us confidence that our analysis has indeed identified a
realistic group formation mechanism.

\begin{figure*}
\includegraphics[width=15.0 cm]{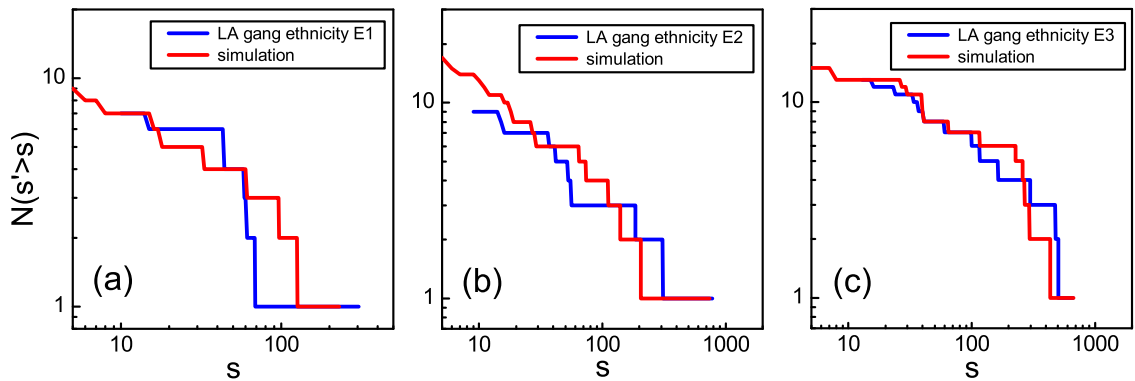}
\caption{(Color Online) The cumulative gang size distribution
$N(s'>s)$ for LA gangs of different ethnicity. (a) $N(s'>s)$ of
membership of LA gangs of ethnicity E1. The parameters used for
the team formation model are $N=608$, $\langle \Delta p_i\rangle =
0.150$, $\sigma_{\Delta p} = 0.016$, and $\tau = 0.73$. (b)
$N(s'>s)$ of membership of LA gangs of ethnicity E2. The
parameters used for the team formation model are $N=1504$,
$\langle \Delta p_i\rangle = 0.142$, $\sigma_{\Delta p} = 0.014$,
and $\tau = 0.72$. (c) $N(s'>s)$ of membership of LA gangs of
ethnicity E3. The parameters used for the team-formation model are
$N=2552$, $\langle \Delta p_i\rangle = 0.141$, $\sigma_{\Delta p}
= 0.016$, and $\tau = 0.72$. Each model result corresponds to one
run of the team-formation model simulation. Note that the
parameters for different ethnic groups are very similar, as was
the case for different servers in WoW.} \label{fig:SIfigureS5}
\end{figure*}

We have tested our model against the empirical data of $N(s'>s)$
for LA gangs data, treating all the gangs collectively
(Fig.~6(e)).  Here, we further test our model for the three major
ethnic groups as shown in Fig.~\ref{fig:SIfigureS3}. Treating the
ethnicity as the analogy of servers in WoW, we counted the number
of members in the gangs in each of the ethnic groups. For the
three major ethnic groups E1, E2, and E3, there are $608$, $1504$,
and $2552$ members, respectively.  These numbers are used as
inputs to our model.  We then adjust the parameter values in the
model to give a distribution $N(s'>s)$ that resembles the
empirical data, for each of the three cases. Figure
\ref{fig:SIfigureS5} shows the results for the ethnic groups E1,
E2, E3.  It is very encouraging that our model manages to capture
the main features of the empirical data for the LA gangs, even
though the individual gang sizes and number of gangs in each
ethnic group are much smaller than for the case of WoW. Note that
the parameters (given in the caption) are quite similar for
different ethnic groups. Interestingly, using the value of $N$
from the empirical data for each of the ethnic groups, the
resulting number of groups in our team-formation model turns out
to be similar to that for the empirical data. From the results of
WoW guilds and street gangs, we can see that the role of server in
WoW has a direct analogy with the role of ethnicity in street
gangs.

In summary, our team-formation model reproduces the main
quantitative features of the empirical WoW guild size distribution
and the cumulative distribution  (Fig.~1(a), Fig.~1(b),
Fig.~\ref{fig:SIfigureS1}), in the case when the servers are
considered collectively (Fig.~6(a)-(d)) {\em and} in the case when
the servers are considered individually
(Fig.~\ref{fig:SIfigureS4}). The model {\em also} reproduces the
main feature in the group dynamics (Fig.~6(a)-(d)) observed in the
empirical data on churn (Fig.1(b) and Fig.~\ref{fig:SIfigureS2}).
Furthermore, the agreement between model and empirical data
extents to results in different time windows (i.e. months).  Our
team-formation model {\em also} reproduces the main quantitative
features of cumulative gang size distributions in empirical data
(Fig.~1(c)), taking the ethnicity collectively (Fig.~6(e)) and
separately (Fig.~\ref{fig:SIfigureS5}). Thus, our self-organized
team-formation model captures quantitatively the features of the
group dynamics resulting from cyber-world interactions, as in the
case of WoW guilds, and real-world interactions as in the case of
street gangs.

\section{V. Inadequacy of the alternative model based on kinship}
There are lines (in light blue) in Fig.~6(b) for WoW server S1 and
in Fig.~6(e) for street gangs that show the results for a kinship
model. The kinship model is in many ways the `opposite' of the
team-formation model, and was introduced to explore homophily as a
possible alternative group-formation mechanism.  In the
team-formation model, the teams tend to recruit members with
attributes that spread over the whole spectrum of attributes,
i.e., the attributes of the agents complement each other.  By
contrast in the kinship model, groups form around agents with
similar attributes. In short, agents tend to dislike being in a
group with agents having very different attributes. Here, we
briefly discuss the mechanisms in this kinship model.

We can readily modify our team-formation model in order to create
a kinship formation model, as follows. The framework in Fig.~5(a)
remains the same, and so does Fig.~5(b) in terms of its structure
-- however we flip the inequalities in Fig.~5(b) for the criteria
for an agent joining a group and for a group accepting a member. A
kinship model can hence be defined which is diametrically opposite
to our team-formation model, and yet can be discussed on the same
footing. In step (a) (see Sec.III), the criteria for an agent $i$
joining a group $J$ are: $|p_{i} - P_{J}| \leq \Delta p_{i}$ and
$f_{i,J} \geq \tau_{J}$. These imply that an agent wants to join a
group with an average attribute close to his own, and a group
wants to accept new members having attributes close to its
existing members.  In step (b)(i), an agent $i$ cannot tolerate a
group $I$ when he finds the members are too different from him.
Thus the agent leaves if $f_{i} < \tau_{i}$.  In step (b)(ii),
each agent is continually looking for a better group which has a
more similar average attribute to him. So group switching happens
if $|p_{i} - P_{J}| < |p_{i} - P_{I}|$ and $f_{i,J} \geq
\tau_{J}$.  Finally, when membership becomes stable, a group tends
to expand by merging with groups having similar average
attributes. Thus two groups $I$ and $J$ merge if $|P_{I} - P_{J}|
\leq min(\Delta P_{I}, \Delta P_{J})$.

In fact, for every team formation model that incorporates the idea
of agents with different attributes tending to form a team, a
corresponding kinship model can be identified, built around the
opposite idea of agents with similar attributes forming groups.
However the cumulative distribution function obtained from the
kinship model cannot capture even the basic qualitative shape of
the empirical data.  The detailed reason is that the kinship model
tends to produce too many groups of small sizes.

\section{VI. Conclusions and implications}
The analysis in this paper contributes to a growing movement
within physics which aims to build quantitative models of
collective dynamics in social systems using the same minimal-model
thinking adopted within physics \cite{Claudio}. We have shown that
populations of humans, in two very different settings, can exhibit
behaviors which are consistent with a common underlying grouping
mechanism. This suggests that many of the collective human
behaviors that we observe, might be driven by common endogenous
features rather than setting-specific exogenous details.

Specifically, we used detailed empirical datasets to show that the
observed dynamics in two very distinct forms of human activity --
one offline activity which is widely considered as a public
threat, and one online activity which is by contrast considered as
relatively harmless -- can be reproduced using the same, simple
model of individuals seeking groups with complementary attributes,
i.e., they want to form a team, as opposed to seeking groups with
similar attributes (homophilic kinship).  Just as different
ethnicities may have different types of gangs in the same city in
terms of their number, size, and stability, the same holds for the
different computer servers on which online players play a given
game.

Our quantitative results provide a novel addition to the group
formation debate by being (i) able to reproduce the quantitative
features of both the dynamical and time-averaged behavior observed
in the empirical datasets, (ii) plausible in terms of the
individual-based rules that are used to describe group membership,
(iii) robust in terms of its insensitivity to small perturbations
in the model's specification and parameter values, (iv) minimal in
that the number of free parameters in the model is kept to a
minimum, given the available datasets to be modelled, and (v) able
to shed light on what mechanistic rules drive people to join and
leave such groups in offline and online situations and provide the
basis for further investigations.

This close relationship that we have uncovered between gangs and
guilds, might be less surprising if it were true that both are
populated by a similar sector of society. However this is not the
case. Online games are played equally by men and women across all
age groups, locations and backgrounds \cite{nic1,nic2,IBM}, while
gangs are mostly populated by teenage urban males from particular
backgrounds \cite{Justice}. Instead, we believe that our results
demonstrate a commonality in the {\em way} in which humans form
such offline and online groups. Interestingly this echoes recent
claims by international law enforcement agencies concerning the
hybrid nature of transnational gangs (`maras'), crime
organizations, insurgencies and terrorist groups, whose
interactions and activities are now beginning to blur the
boundaries between real and virtual spaces \cite{Justice}.

Finally we note that this work throws up the interesting challenge
of providing analytic solutions to accompany the empirical
findings and numerical simulations. Work is proceeding in this
direction, though the difficulty of including internal degrees of
freedom, i.e., the model attributes, in general
coalescence-fragmentation problems is daunting.

\end{document}